\documentclass[referee]{raa}

\usepackage{graphicx,times}
\usepackage{natbib}
\usepackage{amssymb,amsmath}
\bibpunct{(}{)}{;}{a}{}{,}

\usepackage[pagebackref=true]{hyperref}

\begin{document}

   \title{Estimation of Solar Observations with the Five-hundred-meter Aperture Spherical Radio Telescope (FAST)}

 \volnopage{ {\bf 20XX} Vol.\ {\bf X} No. {\bf XX}, 000--000}
   \setcounter{page}{1}

   \author{Lei Qian\inst{1, 2, 3, 4, 5}, Zhichen Pan\inst{1, 2, 3, 4, 5},
   Hongfei Liu\inst{1, 3, 4, 5}, Hengqian Gan\inst{1, 3, 4, 5}, Jinglong Yu\inst{1, 3, 4, 5},
   Lei Zhao\inst{1, 3}, Jiguang Lu\inst{1, 3, 5}, Cun Sun\inst{1, 2, 3},
   Jingye Yan\inst{2, 6}, Peng Jiang\inst{3, 1, 2, 4, 5}}

   \institute{National Astronomical Observatories, Chinese Academy of Sciences,
              20A Datun Road, Chaoyang District,
              Beijing 100101, People's Republic of China; {\it panzc@bao.ac.cn}\\
        \and
              State Key Laboratory of Space Weather, Chinese Academy of Sciences,
              Beijing, 100190\\ 
        \and
             Guizhou Radio Astronomical Observatory, Guizhou University,
             Guiyang 550025, China\\
	    \and
             College of Astronomy and Space Sciences, University of Chinese Academy of Sciences,
             Beijing, 100101, People’s Republic of China\\
        \and
             Key Laboratory of Radio Astronomy, Chinese Academy of Sciences,
             Beijing 100101, People’s Republic of China\\
        \and
            National Space Science Center, Chinese Academy of Sciences,
            Beijing 100190, People’s Republic of China\\
\vs \no
   {\small Received 20XX Month Day; accepted 20XX Month Day}
}

\abstract{
We present the estimation of the solar observation with the Five-hundred-meter Aperture Spherical radio Telescope (FAST).
For both the quite Sun and the Sun with radio bursts,
when pointing directly to the Sun, the total power received by FAST would be out of the safe operational range of the signal chain,
even resulting in the damage to the receiver.
As a conclusion,
the Sun should be kept at least $\sim 2^{\circ}$ away from the main beam during the observing at $\sim 1.25 {\ \rm GHz}$.
The separation for lower frequency should be larger.
For simplicity, the angular separation between the FAST beam and the Sun is suggested to be $\sim 5^{\circ}$ for observations on 200 MHz or higher bands.
}

   \authorrunning{L. Qian et al.}            
   \titlerunning{FAST Solar Observation Estimations}  
   \maketitle

%
\section{Introduction}           
\label{sect:intro}

The solar radio emission is related to the hot plasma and magnetic activities in the solar atmosphere \citep{1985aifo.reptR....J}.
With a brightness temperature of $10^6-10^{12}$ K \citep{1965ASSL....1..342M},
the Sun is one of the strongest and closest radio sources in the sky.

There have been several radio telescopes or arrays with the capability of solar observations around the world \citep{2010LanB...4A..216M}.
Currently, arrays, such as LOFAR \citep{2013A&A...556A...2V}, VLA \citep{1979Natur.278...24L},
ALMA \citep{2018Msngr.171...25B}, and SKA\citep{2019AdSpR..63.1404N}, have their observations of the Sun or plans of the solar-related sciences.
Located at Mingantu, Zhengxiangbaiqi, Inner Mongolia, China,
CSRH (Chinese Spectral RadioHeliograph, \citep{2009EM&P..104...97Y}) or
MUSER (MingantU SpEctral Radioheliograph, the name after its completion) has worked for years.
It is consisted of 40 4.5-m-diameter dishes (covering a frequency range of 0.4-2.0 GHz, with spatial resolutions of 51.6 to 10.3 arcseconds) and 60 2-m-diameter antennas (covering 2-15 GHz, with spatial resolutions of 10.3 to 1.3 arcseconds ) \citep{2021FrASS...8...20Y}.

As the largest single dish radio telescope in the world,
FAST \citep{2011IJMPD..20..989N,2019SCPMA..6259502J,2020RAA....20...64J,2020Innov...100053Q}
is now pushing China's radio astronomy forward to the frontiers in the fields including pulsars, Fast Radio Bursts, and InterStellar Medium.
FAST covers some commonly used bands of solar observations,
e.g., 100-240 MHz for the non-thermal radio solar emission \citep{2018ApJ...852...69S},
and 2.8 GHz which is related to solar extreme ultraviolet emissions on timescales of days and longer \citep{1947Natur.159..405C}.
Nowadays, the FAST 19-beam L-band receiver with a frequency range of 1.05-1.45 GHz is used.
Its beam size is about $3'$, approximately one tenth of the angular diameter of the Sun.
The possible solar observation with this receiver can map the whole disk and atmosphere of the Sun.
With the high sensitivity,
FAST also have the potential to study the short-timescale ($\sim 1\  {\mu} \rm s$ with a channel width of $\sim 1\ \rm MHz$) phenomenon
and track the evolution of the ejected material from the corona to a lower brightness level.

The first and most important thing to consider is the safety of the observation.
In this study, we estimated the power from the Sun received by FAST,
and check if the receiver will be damaged.
The estimation of FAST solar observation was given in Section 2.
The angular separation for proper observation of FAST was discussed in Section 3.
The conclusions were given in Section 4.

\section{Will the quite Sun or Its Bursts Damage FAST?}

Suppose that for a source with brightness temperature of $T_{\rm b}$, the corresponding antenna temperature is $T_{\rm a}$. The power goes into the feed is then
\begin{equation}
    P=kT_{\rm \bf a}\Delta \nu,
\end{equation}
where $k=1.3806\times 10^{-23} {\ \rm J\ K^{-1}}$ is the Boltzmann's constant,
and $\Delta \nu$ is the bandwidth
Typically, for the L-band 19-beam receiver of FAST, $\Delta \nu$ is 400 MHz (1.05 to 1.45 GHz).
In order to work in the linear range,
the threshold of input power for this receiver is roughly estimated as \citep{2021RAA....21..182L},
\begin{equation}
    P_{\rm max}= -70 {\ \rm dBm}=10^{-7} {\ \rm mW} ,
\end{equation}
which can be converted to an antenna temperature threshold as
\begin{equation}
    T_{\rm {\bf a},max,l}=\frac{P_{\rm max}}{k\Delta \nu}\approx 1.8\times 10^4 {\ \rm K},
\label{linear_temperature}
\end{equation}
which depends on the bandwidth.
As another threshold, the input power should be strictly lower than -52 dBm,
or, the low noise amplifier would be damaged.
This limitations corresponds to an antenna temperature of
\begin{equation}
    T_{\rm {\bf a},max,s}\approx 1.1\times 10^6 {\ \rm K}.
\label{safe_temperature}
\end{equation}
Since the angular size of the radio bright region of the Sun is usually larger than the FAST beam,
there is no beam dilution. The antenna temperature equals to the brightness temperature of the Sun.

If the bandwidth of the FAST receivers in the future is between 100 MHz and 1 GHz,
the brightness temperature threshold for damage is between $\sim 4.5\times 10^6 {\ \rm K}$ and $\sim 4.5\times 10^5 {\ \rm K}$.
The brightness temperature of the solar radio emission in the quiescent state can be $\sim 10^6 {\ \rm K}$,
while that of a burst may reach $\sim 10^9 {\ \rm K}$ or higher \citep{1985aifo.reptR....J}.
Obviously, the solar observation with either the current L-band 19-beam receiver or other possible FAST receivers would cause damage.

\section{The Proper Angular Separation of FAST Beam from the Sun}

The radio emmission comes from the upper atmosphere of the Sun. The typical size of the radio bright region of the Sun during an outburst is $\sim 1^{\circ}$ \citep{2021ApJ...906..132C}.

According to the design and laboratory tests \citep{Dunning2017DesignAL},
the power pattern of FAST can be approximated with
\begin{equation}
    P_n(u)=\left[\frac{2^{p+1}p! J_{p+1}(\pi u D/\lambda)}{(\pi u D/\lambda)^{p+1}}\right]^2,
\label{beam_function}
\end{equation}
where $J_{p+1}$ is the Bessel function of the $(p+1)^{th}$ order,
$p=2$ is for the illumination of FAST feeds.
Based on this theoretical beam pattern (see Figure 1), to ensure that the total power received lies in the safe range  (Eq. \ref{safe_temperature}), a bright region of the sun with $T_{\rm b}\sim 10^9 {\ \rm K}$ should be kept $\sim 2$ beams
away from the main beam. To keep FAST working in the linear range (Eq. \ref{linear_temperature}), the bright region should be kept $\sim 5$ beams away. In addition, the Sun should be kept further away to avoid damage from more extreme solar radio bursts, although they may be rare. A 10-beam ($\sim 0.5^{\circ}$, with beam size $\sim1.22\frac{\lambda}{D}$) separation would keep FAST working in the linear range during a $10^{11} {\ \rm K}$ burst.
\begin{figure}[!htp]
   \centering
   \includegraphics[width=12cm]{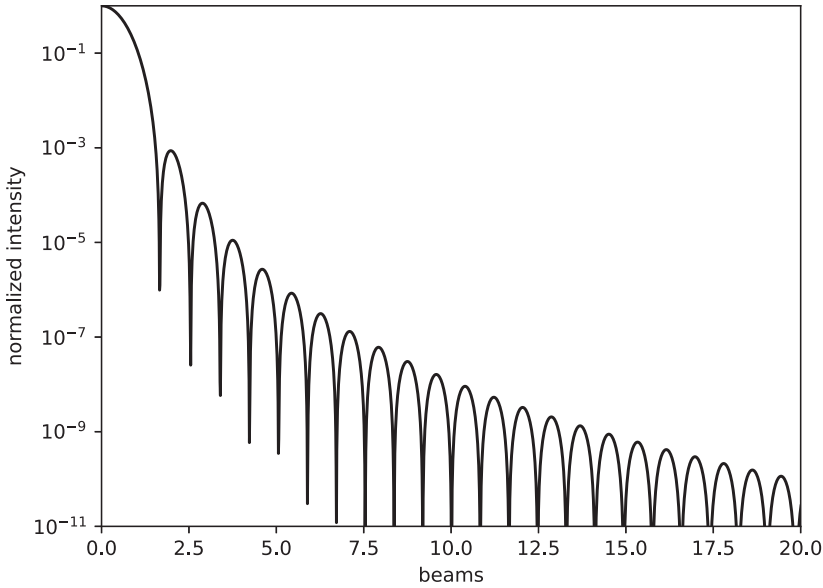}
   \caption{The approximate beam pattern of FAST. }
   \label{beam_shape}
\end{figure}
The diameter of the field of view of the FAST 19-beam receiver is approximately 30'. To summarize, we suggest that the center of the Sun should be kept at least $ 2^{\circ}$ away from the main beam  of the central element of the 19-beam receiver to avoid damage to the receiver.

\begin{figure}[!htp]
   \centering
   \includegraphics[width=12cm]{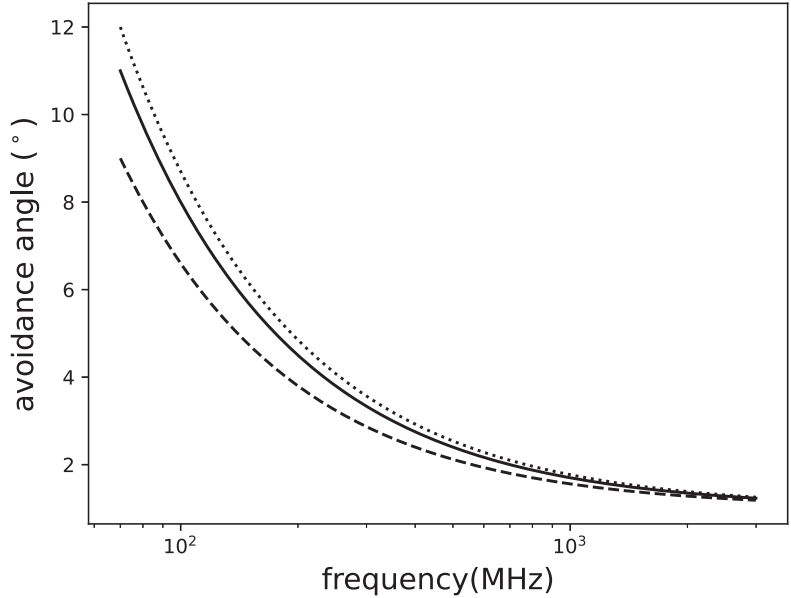}
   \caption{The proper separation between the Sun and the main beam of FAST, as a function of frequency, for a bandwith of 100 MHz (dashed line), 400 MHz (solid line), 1 GHz (dotted line).
            }
   \label{avoid_sun}
\end{figure}

The separation depends on the observing frequency (corresponding to the beam size) and the bandwidth (corresponding to the antenna temperature threshold).
In the designed FAST frequency range (70 to 3000 MHz), it varies from more than 8$^{\circ}$ to around 1$^{\circ}$ (see Figure \ref{avoid_sun})
It is clear that the 5$^{\circ}$ is enough for observation above 200 MHz (the empty circle in Figure \ref{avoid_sun}).

\section{Conclusion}

We have estimated the possibility of FAST solar observation.
The conclusions are as follows:

1. Either the current L-band 19-beam receiver or other possible FAST receivers in the future will be damaged if the telescope directly points to the Sun.

2. Based on the beam pattern and safety considerations,
we suggest that the main beam of FAST should be kept at least $\sim 2^{\circ}$ away from the Sun during the observation when using the L-band 19-beam receiver.

3. The proper angular separation between the FAST beam and the Sun varies with the observing frequency and bandwidth.
For simplicity, a separation of $\sim 5^{\circ}$ for observation above $200 {\ \rm MHz}$ is suggested for most FAST observations.

4. If one considers the solar observation as a science goal of FAST, receivers should be specifically designed to deal with the large power received.

\normalem
\begin{acknowledgements}
This work is supported by National Key R\&D Program of China No.2018YFE0202900, National SKA Program of China No. 2020SKA0120100.
This work is also supported by the Specialized Research Fund for State Key Laboratories, 
National Nature Science Foundation of China (NSFC) under Grant No. 11703047, 11773041, U2031119, 12041303, 12173052, 12003047 and 12173053.
Zhichen Pan is supported by the CAS "Light of West China" Program.
Lei Qian is supported by the Youth Innovation Promotion Association of CAS (id.~2018075), the CAS "Light of West China" Program, and the Science and
Technology Program of Guizhou Province ([2021]4001).
Lei Qian also acknowledge the inspired discussion in the Wechat group "Astrojokes".

\end{acknowledgements}

\bibliographystyle{raa}
\bibliography{bibtex}

\end{document}